\documentclass[twocolumn,pre,aps,showpacs]{revtex4}
\usepackage{graphicx}

\def\be{\begin{equation}}
\def\ee{\end{equation}}
\def\bea{\begin{eqnarray}}
\def\eea{\end{eqnarray}}

\begin{document}

\title{
Quantum corrections and bound-state effects
in the energy relaxation of hot dense Hydrogen. }

\author{M. W. C. Dharma-wardana}
\affiliation{
National Research Council, Ottawa, Canada K1A 0R6}
\email{chandre.dharma-wardana@nrc-cnrc.gc.ca}
\date{\today}

\begin{abstract}

Simple analytic formulae for energy relaxation (ER)
in electron-ion systems, with quantum corrections, ion dynamics
and RPA-type screening are presented. ER in the
presence of bound electrons is examined in view of
of recent simulations
for ER in hydrogen in the range 10$^{20}$-10$^{24}$ electrons/cc.
\end{abstract}

\pacs{52.25.Kn,71.10.-w,52.27.Gr}

\maketitle

{\it Introduction --}
The problem of energy relaxation (ER) in hot electron-ion systems arises in 
astrophysics, fusion physics, condensed-matter physics and
chemistry\cite{murdw,review}.
The large difference in masses $m_e$, $m_i$
between electrons and ions
simplifies the problem somewhat since quasi-equilibrium systems,
 with an electron temperature $T_e$,
and an ion temperature $T_i$ can occur. Such two-temperature plasmas exist in   
material systems as diverse as warm dense matter \cite{Riley}, ultra-cold plasmas 
\cite{Killian}, hot semiconductors \cite{coupledmode1}, and dense deuterium \cite{Knudsen}.
The strong temperature dependence of thermonuclear processes imply that
 estimated burn rates 
depend sensitively on the accuracy of the temperature relaxation (TR) theory used.
Thus a number of recent studies\cite{dwmur1, murdw, mmll}
have examined various aspects of two-temperature dense hydrogen and ER 
in regimes of densities and temperatures relevant to inertial
 confinement fusion (ICF)\cite{ICF}. 

The earliest theories of  ER in  plasmas are due to Landau 
\cite{Land} and Spitzer \cite{Spit} (denoted L-S).  The L-S approach is applicable to 
weakly interacting fully-ionized
plasmas in the classical regime.
It is the Rutherford Coulomb scattering formula applied to Maxwellian
distributions of ions and electrons. Analyses
using the Fermi golden-rule (FGR)
and coupled-mode (CM) extensions
were given by Dharma-wardana et al. \cite{murdw, DWP,mwcd, Hazak}.
Fokker-Plank approaches~\cite{DWP},
kinetic-equation methods based on the relaxation of the
one-particle distributions
\cite{Gould_Dewitt,GMS},
methods based on expansions in the coupling constant\cite{br-sin}
or related techniques\cite{losA},
have been explored in  recent studies\cite{br-sin}.
 Molecular dynamics (MD) was used by Hansen and McDonald,
 (denoted HM) \cite{HM}, using classical potentials
 which incorporate the cutoffs used in the Coulomb logarithm
 of L-S theory.
  The recent studies of Refs.~\cite{mmll,murdw} 
 report more careful applications of the HM approach. 
 The FGR and CM approaches have
 been experimentally tested in semiconductor plasmas,
 but no experimental results of TR 
 are as yet available for hot dense plasmas.

In fully-ionized hydrogen plasma the particle charges $Z_p,Z_e$ are $\pm 1$. 
The mean electron-  and proton densities $n$ and  $\rho$ are identical. 
The ratio of a typical Coulomb energy to the kinetic energy becomes, in the classical 
regime, $\Gamma=1/(r_sT)$, where $T$ is the temperature in energy units, and 
$r_s=\left[3/(4\pi n)\right]^{1/3}$ is the radius of the Wigner-Seitz sphere of an
electron or a proton.
The properties of partially degenerate plasmas
require two independent parameters, e.g., both $\Gamma$ and $\theta=T/E_F$, 
where $E_F=\left(3\pi^2 n\right)^{2/3}/2$ is the
Fermi energy. Thus the regime of densities and temperatures
studied in Ref.~\cite{mmll} involve, at one extreme, 
$r_s=25.25$, $\theta=127$ at $T_e$=10 eV.,
for $n_e=10^{20}$ electrons/cc, while the another extreme is $r_s=1.172$,
 and $\theta=0.274$,
 where $n_e=10^{24}$ electrons/cc at $T_e$=10 eV. The latter is a significantly
 degenerate plasma where the validity of classical-simulation methods is
 suspect. The system at $r_s\sim 25$, $n_e=10^{20}$ electrons/cc 
 at $T_e$ 10 eV contains $1s$ to $3d$ bound states, but the system is essentially
 ionized since the bound-state occupations are negligible.
  On the other hand, the plasma at $n_e=10^{22}$ electrons/cc,
  $T_e=10$ eV, i.e., $r_s$=5.441
 is $\sim 82$\% ionized at $T_e=10$ eV and carries a $1s$ bound state
 at an energy of -0.458 a.u., when calculated using the atom-plasma codes implemented
 by Perrot et al\cite{dpcodes}.  The effect of such bound states cannot be
 included in the classical simulation method of Ref.~\cite{mmll}.

In this communication we use the FGR approach and
derive a simple analytic formula inclusive of leading quantum
corrections, screening and ion-dynamics for for TR in the fully ionized limit, and
compare it with the results from recent classical simulations~\cite{mmll}.
We then consider the effect of bound states, since they can have a significant
effect on energy relaxation.
Many complex processes become possible, but the system with a single
hydrogenic boundstate is a useful case for sharpening our understanding
of this relatively unknown 
regime where no previous results are available.

{\it Quantum transition rates.--}
Assuming that $T_e > T_i$ to be specific, ER occurs via energy transfer from the
excited modes of the electron sub-system to
the cold modes in the ion subsystem.
The spectrum of 
the modes of the species $j$ is given by the spectral
function $A_j(q,\omega,T_j)$.
These spectral functions are given by the imaginary parts of the
corresponding dynamic response functions $\chi^{j}(\vec{k},\omega)$, e.g,
Eq.~(16) of Ref.~\cite{Hazak}.
 The ER rate evaluated within the Fermi golden rule, $R_{fgr}$
 can be expressed in terms of 
the response functions of the plasma as given in Eqs. (4)-(7)
 of Ref.\cite{mwcd}, and Eq.~(15) of Ref.~\cite{Hazak}:
\bea
\label{cothform}
R_{fgr}&=&\frac{\delta E}{\delta t}=\int \frac{d^{3}k}{\left( 2\pi \right) ^{3}} 
\frac{\omega d\omega}
{2\pi}(\Delta B) F_{ep}\\
\Delta B&=&\coth(\omega/2T_e)-\coth(\omega/2T_p)\\
F_{ep}&= &|( V_{ep}(k)
| ^{2}\Im\left[ \chi ^{p}(\vec{k},\omega )\right] \Im
\left[ \chi ^{e}\left( \vec{k},\omega \right) \right]
\eea
In the above $\delta E/\delta t$ is the rate of change of the energy
of the system, for time steps $\delta t$ significantly greater than the
equilibriations times $\tau_p, \tau_p$ which establish $T_e$ and $T_p$
of each subsystem. The relaxation of the whole system is determined
by $\tau_{ep}$ such that $\tau_{ep}>>\tau_p>\tau_e$. 
For brevity we write $\delta E/\delta t$  as $dE/dt$.
we have used the spherical symmetry of the plasma
 to write scalars $q,k$ instead of
$\vec{q},\, \vec{q} $ to simplify the notation.
The  non-interacting response function $\chi_0(q,\omega,T)$ at arbitrary
degeneracies was given by Khanna and Glyde\cite{KH}, and are used here
 in the generalized RPA $\chi^j(q,\omega,T)$ form, $j =e,p$
where, for example, temperature-dependent
local field corrections $G_{ee}(k)$ may be included\cite{prb3d}. 
If $T_e,T_i$ are {\em both} sufficiently large that
$\Delta B\to 2(T_e-T_i)/\omega$, and if the electron
 chemical potential $\mu_e \le 0$,
useful analytical approximations become available. The possibility
 of extracting a temperature-relaxation time
$\tau_{ep}$ from the relaxation rate exists only in this
regime.  Neglecting
interactions, $E$ becomes the kinetic energy. Using non-interacting classical forms
for $\Im\chi_0^j(k,\omega)$ in Eq.~\ref{cothform} we obtain the
well known Landau-Spitzer (L-S) form
for the temperature relaxation time $\tau$, viz., 
\bea
1/\tau&=&\frac{2}{3n}\omega_{p_e}^2\omega_{p_p}^2\left[(2\pi T_{ep})/m_{ep}\right]^{-3/2}{\cal L}\\
 {\cal L} &=& \log(k_{max}/k_{min}) \\
  T_{ep}/m_{ep}&=&T_e/m_e+T_p/M_P, \;\;\; \omega_{p_j}^2=4\pi n/m_j
\label{coullog}
\eea
Here $\omega_{p_j}$ is the plasma frequency of the  species $j =e,p$.
 The effective temperature
and the effective mass of the colliding pair are $T_{ep}$ and $m_{ep}$,
  with $T_j$ in energy units.
 ${\cal L}$ is the
``Coulomb logarithm''. It depends on $k_{min}$ and $k_{max}$, i.e., 
  momentum cutoffs (or impact parameters) used for modeling the
  unscreened Coulomb 
collision. If interacting response functions 
(e.g, RPA) are used, single-particle modes become replaced almost
completely by plasmon modes, and the interactions become
dynamically screened. Any type of ``static screening'' must satisfy the $f$-sum
rule if false results are to be avoided. 
\begin{figure}[t]
\includegraphics[angle=0, width=3.3in]{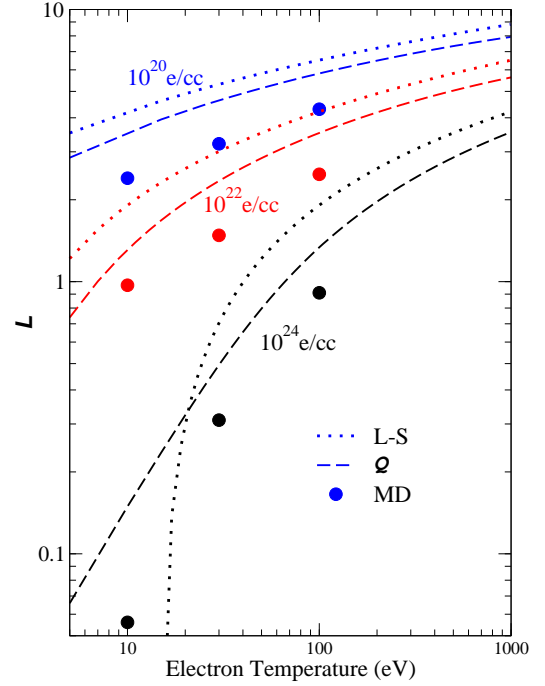}
\caption{The Landau-Spitzer Coulomb Logarithm (${\cal L}$) is
compared with the classical MD simulations\cite{mmll} and
the analytic Coulomb factor ${\cal Q}$, Eq.~\ref{newresult1}
derived from the FGR result.
}
\label{taucomp1}
\end{figure}
 The safest procedure is to do the $\omega$-integration
imposing the $f$-sum rule\cite{Hazak}. Then ion dynamics are automatically
preserved,
and Eq.~\ref{cothform} simplifies to:
\bea
\frac{1}{\Delta}\frac{d\Delta}{dt}&=&\frac{2}{3n}\omega _{p_p}^{2}\int_0^\infty\frac{2}{\pi} 
\left[ \frac{\partial }{\partial \omega }\Im\chi ^{ee}\left(
k,\omega \right) \right] _{\omega =0}dk
\label{fgrBA}
\eea
We approximate  $\Im\partial\chi^{ee}/\partial\omega\,|_{\omega =0}$  as:
\bea
\Im \partial\chi^{ee}/\partial\omega\;|_{\omega=0}&=&\frac{\Im\partial\chi_0^{ee}/
\partial\omega\;|_{\omega=0}}
{\{1+k^2_{sc}/k^2\}^2}
\label{derivchi}
\eea
The anti-symmetry of the imaginary part has been used in the above.
 The electron-screening wavevector
$k_{sc}$ at any degeneracy is obtained from the small-$q$ limit of the
finite-$T$ Lindhard function. The $k\to 0$-local field correction, $G_0^{ee}$ at
arbitrary degeneracy\cite{prb3d} can
also be included in $k_{sc}$ {\it via} the following definitions.
\bea
(k^0_{sc})^2&=&\frac{2}{\pi}(2T)^{1/2}I_{-1/2}(\mu^0_e/T_e)\\
I_{\nu}(x)&=&\int_0^{\infty}\frac{dyy^\nu}{e^{y-x}+1},\, \nu\ge-\frac{1}{2} \\
k_{sc}&=&k^0_{sc}\left[1-G_0^{ee}\right]^{1/2}
\label{screenkc}
\eea 
However, in this study we find it sufficient to use he Debye-H\u{u}kel form $k_{sc}=k^e_{DH}$
for electron screening.
In approximating $\Im\partial\chi^{ee}/\partial\omega$ we retain
terms up to second order in $\hbar$ as displayed explicitly below:
\be
\Im\chi_0 ^{ee}=-(\frac{\pi}{2T_e})^{3/2}\,\frac{2n\omega}{\pi k}
e^{-\frac{1}{2T_e}\{\frac{\omega^2}{k^2}-\frac{\hbar^2k^2}{4}\}}
\frac{\sinh(\hbar\omega/2T_e)}{\hbar\omega/2T_e}
\ee
Then Eq.~\ref{fgrBA} can be reduced to the form:
\bea
1/\tau&=&-\frac{2}{3n}\omega_{p_e}^2\omega_{p_p}^2\{2\pi(T_{ep}/m_{ep})\}^{-3/2}{\cal Q}\\
{\cal Q}&=&\frac{1}{2}\left[e^pEi(p_e)(p_e+1)-1\right]\\
p_e&=&k_{sc}^2/(8T_e)\\
Ei(x)&=&\int^{\infty}_x exp(-t)dt/t
\label{newresult1}
\eea
The exponential integral~\cite{gradstein}, $Ei(x)$ of Eq.~\ref{newresult1},
is evaluated numerically via standard subroutines.
Thus we see that the ``Coulomb factor'' ${\cal Q}$ 
is exactly analogous to the ``Coulomb logarithm'' of
Eq.~\ref{coullog}, but without {\it ad hoc} cutoffs.
${\cal Q}$ contains leading-order quantum corrections,
 ion-dynamics and electron
screening. The expression for ${\cal Q}$ should be compared with a
similar expression given by Brown et al\cite{br-sin}.
\be
{\cal L}_{bps}=\frac{1}{2}\left[\log(1/p_e)-\gamma-1\right]
\ee
where $\gamma$=0.5772 is the Euler constant. At high $T_e$, this result
approaches the L-S form more rapidly than ${\cal Q}$. 

\begin{figure}[t]
\includegraphics[angle=0, width=3.3in]{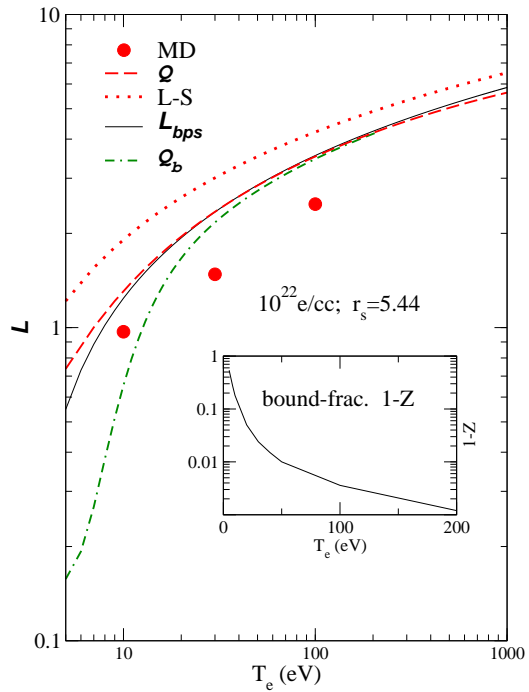}
\caption{The effect of bound states in the H-plasma with
$r_s=5.44$, shown as ${\cal Q}_b$, 
compared with calculations without bound-state 
effects. ${\cal L}_{bps}$ is also shown to closely
approximate ${\cal Q}$. 
The bound fraction $1-Z$, where $Z$ is the degree of
ionization, is shown in the inset.
}
\label{taubou}
\end{figure}

The interactions between the ion- and electron modes lead
to ion-acoustic modes (coupled modes)~\cite{murdw}. 
The H-plasmas treated here may be considered
relatively weekly coupled plasmas, and the correction from
coupled-mode effects will
be neglected. The results shown in Fig.~\ref{taucomp1} suggest
that the classical potentials etc., used for $r_s=5.44$ and $r_s=25.25$
need reconsideration, since these are weakly coupled
plasmas where FGR methods should give good agreement.
As seen from
Fig.~\ref{taubou}, the numerical results from ${\cal Q}$
 and ${\cal L}_{bps}$ very similar.
 The case $r_s\simeq 1$ has been discussed
in more detail elsewhere\cite{murdw}.

{\it Energy Relaxation in the presence of bound states--}
The nuclei in the  H-plasma with $10^{22}$ electrons/cc, $r_s=5.44$ carry a single $1s$ bound
state of energy $\sim$0.46 a.u. with the effective charge $Z$ varying from 0.82
at $T_e$=10 eV, to $\sim 1.0$ at $T_e>200$ eV. If we use an ``average ion'' picture,
we have a gas of ions with change $Z$, a boundstate occupation $n_{1s}=1-Z$ electrons
per ion, and the remaining (``free'') electrons distributed in continuum states.
 The electrons in the bound states
have an effective mass $m_{ep}\simeq m_e=1$, and hence equilibrate rapidly with
the hot electrons in the continuum. That is, the distribution $n_{1s}$, and hence
the degree of ionization $Z$ is determined by the hot electron temperature $T_e$.
(See fig.~\ref{taubou} for a plot of $1-Z$). 
The center of mass (CM)  motion of the ions and bound electrons is determined
 by the kinetic energy of the mass $M_{CM}=M_p+m_e\simeq M_P$, i.e., CM motion is at the
 temperature $T_p$. The energy exchanging collisions are between the hot electrons
 and the ions. The bound electrons (in their $1s$ states) do not interact with their own
 binding nuclei. The energy mismatch in the spectral functions ensures that they do not
 significantly interact with other ions carrying bound states. This ``average-ion''
picture holds if the temperature relaxation time $\tau$ is sufficiently long compared
to the ionization-equlibriation time $\tau_Z$. This picture is applicable to
the system discussed in this study, and the only source of energy relaxation is via Coulomb
collisions.

An alternative picture holds if $\tau$ is comparable to  $\tau_Z$.
 Then the plasma consists of
nonionized atomic hydrogen with fully occupied $1s$ states at some temperature $T_a$
of the atoms, free electrons at
$T_e$ and protons at $T_p$. The atom temperature $T_a$ has to be self
consistently determined
via a theory of ionizing collisions as well as the Coulomb collisions
discussed so far.
This regime occurs for low $T_e, T_p$ and is not considered here.

The energy $E_e$ of the electron subsystem, and the energy $E_i$ of the CM-ion subsystem
in the average-ion model with hydrogenic states $\epsilon_\nu$, occupations $n_v$ 
($\nu=n,l,m$ being the set of quantum numbers of the state) is given by:
\bea
E_I&=&3nT_p/2+ E^I_{int}\\
E_e&=&3nZT_e/2+n(1-Z)\sum_\nu(\epsilon_\nu-\mu_a)n_\nu   \\
   &\simeq&3nZT_e/2+n(1-Z)^2\epsilon_{1s}+ E^e_{int}
\label{ebound}   
\eea
The terms $E^J_{int}$ involve contributions from particle interactions, embedding
effects, continuum lowering effects etc., which have been discussed at length in
Ref.~\cite{multiconf}. These effects can be neglected for the weekly-coupled 
 H-plasmas studied here.
The last equation limits the problem to one bound state.
The effects arising from the chemical potential $\mu_a=\mu_e(T_e)+\mu_p(T_p)+\delta\mu_a$ will
be neglected in all the species, free and bound, as they are found to be small. Then the time
dependence $dE_j/dt$ can be replace by $dE_j/dT_j)(dT_j/dt)$ and we attempt to construct a
formula for temperature relaxation. Unlike in simple classical
 plasmas, it is no longer possible to define a
temperature relaxation time $\tau$. Hence, to be specific we may thermostat $T_i$
 and consider the relaxation rate $dT_e/dt$,
with ($Z\ne 1$)  and without ($Z=1$)  bound-state effects. 
Thus we have:
\bea
dT_e/dt|_{Z=1}&=&(T_e-T_i)C_{pe}{\cal Q}(p_e)\\
C_{pe}&=&\frac{2}{3n}\omega_{p_e}^2\omega_{p_p}^2\{2\pi T_{ep}/m_{pe}\}^{-3/2} \\
dT_e/dt|_{Z\ne1}&=&(T_e-T_i)\frac{Z^3C_{pe}}{B_{pe}}{\cal Q}(Zp_e)\\
1/B_{pe}&=&\left[Z+\{T_e-\frac{4}{3}(1-Z)\epsilon_{1s}\}\frac{dZ}{dT_e}\right]\nonumber
\label{dTebound} 
\eea
This result shows that the effective ionization $Z$ enters approximately as $Z^2$ in the
prefactor. The dependence of the ionization temperature, viz., $dZ/dT_e$ was obtained from
a DFT calculation\cite{dpcodes}. It may be argued that some of these corrections
are more part of the specific heat of the electron system  inclusive of
bound states, and do not belong to the collision dynamics.
However, merely to compare the $Z=1, Z\ne1$ systems,
we define
\be
{\cal Q}_b= \{Z^3/B_{pe}\}{\cal Q}(Zp_e)
\ee
We display ${\cal Q}$ and ${\cal Q}_b$ using the
Eqs.~\ref{dTebound}, in Fig.~\ref{taubou}, together with
the classical MD, ${\cal L}_{bps}$ and L-S results which treat only
the fully ionized state. The inset
to the figure shows the degree of association ($1-Z$) as
a function of temperature, calculated using density-functional
methods\cite{dpcodes}.

{\it Conclusion.--}We presented analytic formulae with quantum corrections,
ion dynamics and screening .   They have been
applied to weakly coupled H-plasmas ($r_s=5.44$, 25.25) where the physical approximations
are felt to be reliable. The differences between the classical MD simulations
and our results suggest the need for further review of the classical 
simulations as well as the physical theory.
The effect of the bounds states on a quantity nominally similar
to the Coulomb logarithm has also been presented for the H-plasma 
at the density 10$^{22}$ e/cc. This too is an area which 
requires further study and poses a novel challenge to simulations..




\begin{thebibliography}{99}
\bibitem{murdw} M. S. Murillo and M. W. C. Dharma-wardana, Phys. Rev. Lett. (submitted)
\bibitem{review}M. W. C. Dharma-wardana, ``Density functional and non-equilibrium methods
for unusual states of matter produced using short-pulse lasers'',
in {\it Laser Interactions with Atoms, Solids, and Plasmas} (NATO ASI, series B
vol.327), p 331 Edited by R.M.More (Plenum, N.Y) 1994. 
\bibitem{Riley} D. Riley, N. C. Woolsey, D. McSherry, I. Weaver, A. Djaoui, and E. Nardi,
{\it Phys. Rev. Lett.} {\bf 84}, 1704 (2000).
\bibitem{Killian} Y. C. Chen, C. E. Simien, S. Laha, P. Gupta, Y. N. Martinez, P. G. Mickelson,
S. B. Nagel, T. C. Killian, {\it Phys. Rev. Lett.} {\bf 93}, 265003 (2004).
\bibitem{coupledmode1}M.W.C. Dharma-wardana,  {\it Phys. Rev. Lett.} {\bf 66}, 197-201 (1991); 
P. Celliers, A. Ng, G. Xu, and A. Forsman, {\it Phys. Rev. Lett.} {\bf 68}, 2305 (1992).
\bibitem{Knudsen} M. D. Knudson, D. L. Hanson, J. E. Bailey, C. A. Hall, and J. R. Asay,
{\it Phys. Rev. Lett.} {\bf 90}, 035505 (2003).
\bibitem{dwmur1} M. W. C. Dharma-wardana and M. S. Murillo, Phys. Rev. E.  {\it 77}, 026401 (2008)

\bibitem{mmll}J. N. Glosli, F. R. Graziani, R. M. More, M. S. Murillo, F. H. Streitz, M. P. Surh, L. X. Benedict,
 S. Hau-Riege, A. B. Langdon, R. A. London, http://arxiv.org/pdf/0802.4037 
\bibitem{ICF} R. Linford, R. Betti, J. Dahlburg, J. Asay, M. Campbell, Ph. Colella, J. Freidberg, J. 
Goodman, D. Hammer, J. Hoagland, S. Jardin, J. Lindl, G. Logan, K. Matzen, G. Navratil, A. Nobile, 
J. Sethian, J. Sheffield, M. Tillack and J. Weisheit, {\it J. Fusion Energy} {\bf 22}, 93 (2003).
\bibitem{Land} L. D. Landau, JETP {\bf 7}, 203 (137); E. M. Lifshitz and L. P. Pitaevskii, {\sf Physical 
Kinetics} (Pergamon, Oxford, 1981).
\bibitem{Spit}  L.\ Spitzer, {\sf Physics of Fully Ionized Gases} (Interscience, New York, 1967).
\bibitem{DWP} M. W. C. Dharma-wardana and F. Perrot, {\it Phys. Rev. E} {\bf 58}, 3705 (1998);
{\it Phys. Rev. E} {\bf 63}, 069901 (2001).
\bibitem{mwcd}
M. W. C. Dharma-wardana, Phys. Rev. E {\bf 64} 035401 (2001)
\bibitem{Hazak} G. Hazak, Z. Zinamon, Y. Rosenfeld, and M. W. C. Dharma-wardana, {\it Phys.
Rev. E} {\bf 64}, 066411 (2001).
\bibitem{Gould_Dewitt} H. A. Gould and H. E. DeWitt, {\it Phys. Rev.} {\bf 155}, 68 (1967).
\bibitem{GMS} D. O. Gericke, M. S. Murillo, and M. Schlanges, {\it Phys. Rev. E} {\bf 65}, 036418 
(2002).
\bibitem{br-sin}L. S. Brown, D. L. Preston and R. L. Singleton, Jr., Phys. Rep. {\bf 410}, 237 (2005)
\bibitem{losA} J\`{e}r\^{o}me Daligault, Dmitry Mozyrsky, Phys. Rev. E {\bf 75}, 026402 (2007);
We do not agree with the claim made by these authors that the coupled-mode formulations do not
reduce to the binary-collision result at sufficiently high temperatures.
\bibitem{HM} J. P. Hansen and I. R. McDonald, {\it Phys. Lett.} {\bf 97A}, 42 (1983).
\bibitem{dpcodes}http://athens.phy.nrc.ca/ims/qp/codes/chandre/D\_P/
These codes may be accessed and run via the internet, using a password available from the author.
\bibitem{KH}
F. C. Khanna and H. R. Glyde, Can .J. Phys. {\bf 54}, 648 (1978);
\bibitem{prb3d}
F. Perrot and M. W. C. Dharma-wardana, Phys. Rev. B15  {\bf 62}, 16536 (2000)
Erratum {\bf 67}, 79901 (2003)
\bibitem{gradstein}
I. S. Gradstein, I. M. Ryzhik, {\em Tables of Integrals, Series and products},
 (Academic, New York 1980) Section 8.2
 \bibitem{multiconf}
 F. Perrot and M. W. C. Dharma-wardana, Phys. Rev. E {\bf 52} 5352 (1995) 

\end{thebibliography}
\end{document}